\documentclass[submission,copyright,creativecommons]{eptcs}
\providecommand{\event}{HAS 2013} % Name of the event you are submitting to
\usepackage{breakurl}             % Not needed if you use pdflatex only.
\usepackage{graphicx,psfrag,epsfig,epsf}

\title{Networked Embedded Control Systems: \\ from Modelling to Implementation}
\author{Maria Domenica Di Benedetto
\institute{Department of Information Engineering, Computer Science and Mathematics, \\
Center of Excellence DEWS,\\
University of L{'}Aquila, 67100 L{'}Aquila, Italy.}\\
\email{mariadomenica.dibenedetto@univaq.it}
\and
Giordano Pola
\institute{Department of Information Engineering, Computer Science and Mathematics, \\
Center of Excellence DEWS,\\
University of L{'}Aquila, 67100 L{'}Aquila, Italy.}\\
\email{\quad giordano.pola@univaq.it}
}
\def\titlerunning{Networked Embedded Control Systems: from Modelling to Implementation}
\def\authorrunning{M.D. Di Benedetto, C. Author \& G. Pola}

\begin{document}
\maketitle

%\begin{abstract}
%This is a sentence in the abstract.
%This is another sentence in the abstract.
%This is yet another sentence in the abstract.
%This is the final sentence in the abstract.
%\end{abstract}

%\section{Introduction}

\begin{center}
\textbf{Extended Abstract}
\end{center}

\begin{figure}
\begin{center}
\includegraphics[scale=2]{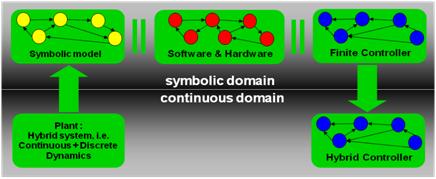}
\caption{Correct-by-Design Embedded Control Software.}
\label{fig1}
\end{center}
\end{figure}

Discrete abstractions of continuous and hybrid systems have been the topic of intensive study in the last twenty years from both the control systems and the computer science communities, see e.g. \cite{TACsymbolicmodels}. While physical world processes are often described by differential equations, digital controllers and software and hardware at the implementation layer are usually modeled through discrete/symbolic processes. These mathematical models heterogeneity has posed during the years interesting and challenging theoretical problems that must be addressed in order to ensure the formal correctness of control algorithms in the presence of non-idealities at the implementation layer. \\
From the synergistic collaboration of researchers in the control systems and computer science communities a novel and sound approach has recently emerged, which is termed "Correct-by-Design Embedded Control Software". This research line can be roughly described as a three-step process, as shown in Figure \ref{fig1}, and detailed hereafter:

\begin{itemize}
\item A finite state machine (or symbolic model) is constructed, which is equivalent or approximates the continuous control system.
\item The original control design problem is solved at the discrete abstraction layer on the symbolic model.
\item The symbolic controller synthesized at the discrete layer, is appropriately refined so that it can be applied to the original continuous control system.
\end{itemize}

The correct-by-design approach guarantees that controllers synthesized at the symbolic layer satisfy the desired specification on the continuous layer, with guaranteed approximation bounds. Moreover, this approach provides the designer with a systematic method for addressing a wide spectrum of novel specifications that are difficult to enforce by means of conventional control design, for example logic specifications expressed in linear temporal logic or automata on infinite strings. \\

\begin{figure}
\begin{center}
\includegraphics[scale=0.4]{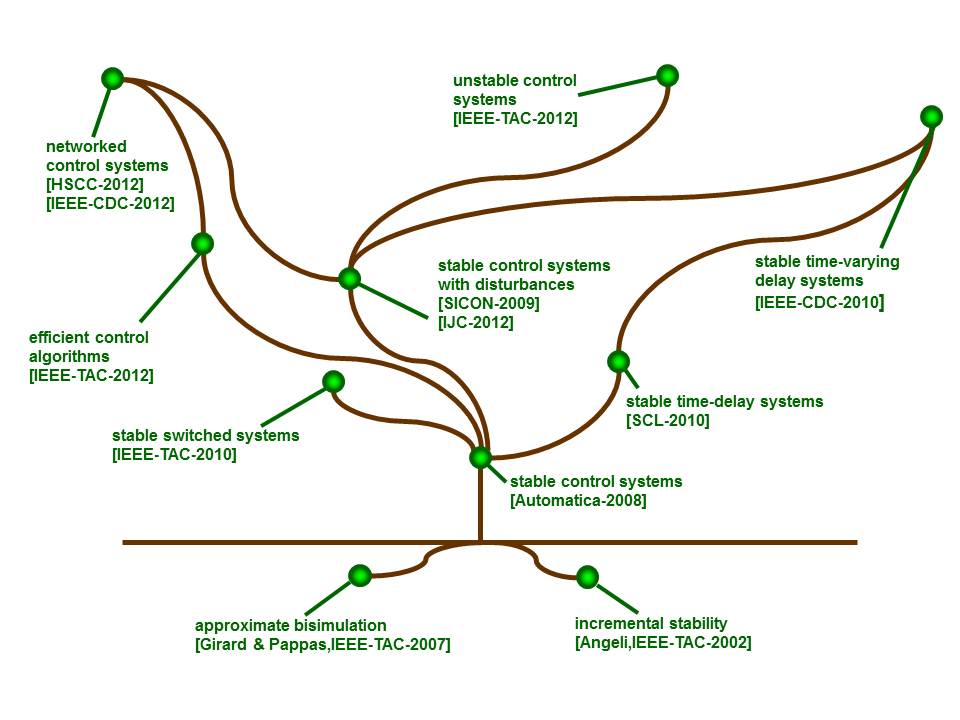}
\caption{Directed Acyclic Graph summarizing research activities of DEWS on Correct-by-Design Embedded Control Software.}
\label{fig2}
\end{center}
\end{figure}

The core of this approach is the definition and construction of symbolic models that are equivalent or approximate continuous and hybrid systems (Step 1 in the methodology).
Several classes of dynamical and control systems that admit equivalent symbolic models have been identified in the literature: for example, timed automata \cite{alur}, rectangular hybrid automata \cite{puri}, and o-minimal hybrid systems \cite{lafferriere,brihaye}. The notion of bisimulation \cite{Milner,Park} is an essential ingredient to capture equivalence between dynamical, hybrid  \textit{infinite state} systems and the corresponding symbolic models. Early results for classes of control systems were based on dynamical consistency properties \cite{caines}, natural invariants of the control system \cite{koutsoukos}, $l$-complete approximations \cite{moor}, and quantized inputs and states \cite{forstner,BMP02}. Recent results include work on controllable discrete-time linear systems \cite{TabuadaLTL}, piecewise-affine and multi-affine systems \cite{habets,BH06}, set-oriented discretization  for discrete-time nonlinear optimal control problems \cite{junge1} and abstractions based on convexity of reachable sets \cite{gunther}. \\
%The notion of approximate bisimulation, a generalization of the notion of bisimulation to metric systems, introduced by Girard and Pappas \cite{AB-TAC07}, provided fruitful insights into this research line and inspired research to identify other classes of control systems admitting symbolic models, examples of which are incrementally stable and incrementally forward complete nonlinear control systems with and without disturbances \cite{PolaAutom2008,MajidTAC11,PolaSIAM2009,BorriIJC2012}, switched systems \cite{GirardTAC2010} and time-delay systems \cite{PolaSCL10,PolaCDC10}. The interested reader is referred to \cite{GirardEJC11,paulo} for an overview on recent advances in this domain. \\
DEWS researchers have been active in this research topic for years. Their collaboration with researchers from the University of California at Los Angeles (USA) and the Universit\'e Joseph Fourier (France) has been fruitful, as demonstrated by the publications summarized in Figure \ref{fig2}. \\
The initial goal was developing a theory towards the definition and construction of symbolic models for nonlinear control systems. We identified two key ingredients to accomplish this ambitious goal: the notion of approximate bisimulation, introduced by Antoine Girard and George Pappas in \cite{AB-TAC07}, and the one of incremental input-to-state stability introduced by David Angeli in \cite{IncrementalS}. We showed in \cite{PolaAutom2008} that these notions can be combined so that for any incrementally input-to-state stable nonlinear control system with compact state space, constructing a symbolic model is possible that approximates the original system with arbitrarily good accuracy in the sense of approximate bisimulation. This result was then generalized to more general classes of continuous and hybrid systems, as follows. \\
First, nonlinear control systems affected by disturbances were considered. In this context an appropriate notion of approximate equivalence, alternating approximate bisimulation, introduced in \cite{PolaSIAM2009} has been obtained by combining approximate bisimulation introduced by Antoine Girard and George Pappas in \cite{AB-TAC07} with alternating bisimulation introduced by Rajeev Alur and co-workers in \cite{Alternating}. This notion guarantees that control strategies synthesized on symbolic models, based on alternating approximate bisimulations, can be readily transferred to the original model, independently of the particular evolution of the disturbance inputs. In this regard, we showed that for any incrementally globally stable nonlinear control system affected by disturbances it is possible to construct a symbolic model which approximates the original system with arbitrarily good accuracy in the sense of alternating approximate bisimulation. \\
The results obtained for nonlinear control systems were then generalized in \cite{GirardTAC2010} to switched systems, a class of hybrid control systems. After generalizing the theory of incremental stability from nonlinear control systems to nonlinear switched systems we proposed symbolic models that approximate incrementally globally asymptotically stable nonlinear switched systems with arbitrarily good accuracy. The proposed theory was then tested on the Boost DC-DC converter, a benchmark selected within the Network of Excellence HYCON consortium \cite{Hycon}.\\
Our next research goal was to investigate further results which can ensure formal correctness of control algorithms in the presence of delays which are rather frequent in the exchange of information in distributed embedded systems. We therefore faced the problem of generalizing the results concerning the construction of symbolic models from the class of nonlinear control systems to the one of nonlinear time-delay systems \cite{PolaSCL10}. This generalization was not straightforward because time-delay systems are infinite dimensional systems. We first generalized the notion of incremental input-to-state stability from nonlinear control systems to nonlinear time-delay systems and provide sufficient conditions for checking this property  in terms of Lyapunov-Krasovskii types dissipation inequalities. We then proposed first order spline-based approximation schemes to approximate the state of the time-delay system. The result achieved in this context mimic the one of nonlinear control systems: any incrementally input-to-state stable time-delay system admits a symbolic model which is approximately bisimilar to the original system. \\
All these results were based on a notion of incremental stability. In collaboration with researchers from the University of California at Los Angeles, we recently relaxed this assumption and showed in \cite{MajidTAC11} that any incrementally forward complete nonlinear control system admits symbolic models that approximate the original system in the sense of alternating approximate simulation. The incremental forward completeness assumption is a rather mild assumption which is fulfilled, for example, by unstable linear control systems. \\
The use of symbolic models for the control design of continuous and hybrid systems provides the designer with a systematic method to address a wide spectrum of novel specifications that are difficult to enforce by means of conventional control design paradigms. In this context we faced the problem of designing symbolic controllers that enforce a nonlinear control system to satisfy a specification expressed in terms of automata theory on infinite strings and so that the interaction between the nonlinear control system and the symbolic controller is non-blocking. An explicit solution to this problem has been derived in \cite{PolaTAC12}, resulting in the non-blocking part of the approximate parallel composition between the specification automaton and the symbolic model of the continuous system. Efficient control algorithms were also derived which cope with the inherent complexity of the approach.\\
All aforementioned results do not consider non--idealities at the communication infrastructure conveying information from the process to the controller and vice versa. We exploited this problem and we derived conditions for the construction of symbolic models of incrementally stable nonlinear control systems in \cite{BorriHSCC12} and for incrementally forward complete nonlinear control systems in \cite{BorriCDC2012}. In particular, we considered  relevant non--idealities arising in a real communication network such as quantization errors, time-varying delay in accessing the network, time-varying delay in delivering messages through the network, limited bandwidth and packet dropouts.

%\nocite{*}
\bibliographystyle{eptcs}
\bibliography{biblio2}
\end{document}